\documentclass[fluids,article,submit,moreauthors,pdftex,10pt,a4paper]{mdpi} 

\Title{Rheology of an inverted cholesteric droplet under shear flow}

\Author{F. Fadda $^{1,2}$, G. Gonnella $^{2}$,  A. Lamura $^{3}$, E. Orlandini $^{4}$ and A. Tiribocchi $^{4,5}$}

\address{%
$^{1}$ \quad Department of Chemical Engineering, Kyoto University, Kyoto 615-8510, Japan; fede.fadda1110@gmail.com\\
$^{2}$ \quad Department of Physics and Sezione INFN Bari, 70126 Bari, Italy; gonnella@ba.infn.it\\
$^{3}$ \quad Istituto Applicazioni Calcolo, CNR, 70126 Bari, Italy; a.lamura@ba.iac.cnr.it\\
$^{4}$ \quad Departmento of Physics and Astronomy and Sezione INFN, Università di Padova, 35131 Padova, Italy; orlandini@pd.infn.it\\
$^{5}$ \quad Center for Life Nano Science, IIT, 00161 Roma, Italy; adriano.tiribocchi@iit.it}

\corres{Correspondence: adriano.tiribocchi@iit.it}

\abstract{The dynamics of a quasi two-dimensional isotropic droplet in a cholesteric liquid crystal medium under symmetric shear flow is studied by lattice Boltzmann simulations. We consider a geometry in which the flow direction is along the axis of the cholesteric, as this setup  exhibits a significant viscoelastic response to external stress. We find that the dynamics depends upon the magnitude of the shear rate, the anchoring strength of the liquid crystal at the droplet interface and the chirality. While for low shear rate and weak interface anchoring the system  shows a non-Newtonian behavior, a Newtonian-like response is observed at high shear rate and strong interface anchoring. This is investigated both by estimating the secondary flow profile, namely a flow emerging along the out-of-plane direction (absent in  fully Newtonian fluids, such as water), and by monitoring defect formation and dynamics which alter significantly the rheological response of the system.}


\begin{document}

\section{Introduction}

Liquid crystals (LC) are fluids often made up of rod-like molecules that can arrange in a variety of equilibrium phases, depending upon their geometry and reciprocal interaction~\cite{deGennes,Chandra}. Of particular relevance is the cholesteric  phase where the local molecular alignment is captured by a unit magnitude director field ${\bf n}$ displaying a natural twist deformation in the direction perpendicular to the molecules. Such liquid crystals can be found in a wide range of biological systems, such as bacterial flagella~\cite{berg} and DNA molecules in solution~\cite{watson}, and have found increasing application in modern display devices, in which the cholesteric phase results from mixing a nematic liquid crystal (such as E7) with a chiral dopant. In recent years much interest has been focused on the realization of liquid crystal - particles composites, a new exciting class of soft material with tunable elastic and electro-optic properties, in which either colloidal particles or droplets of standard fluids are dispersed in a LC phase~\cite{musevic}. The presence of these inclusions induces a disturbance in the director orientation which typically aligns either tangentially or normally to the surface of the particles. This leads to the formation of a variety of topological defects that mediate long-range anisotropic particle-particle interactions and stabilize novel ordered structures, such as chain~\cite{loudet,poulin} and defect glass~\cite{wood} in nematics or colloidal crystals~\cite{ravnik} and isolated clusters~\cite{lintuvuori} in cholesterics, with potential use in photonics~\cite{smalyukh} and in new devices~\cite{chari}. 

Recent theoretical and experimental works have investigated the typical defect structures observed at equilibrium when particles are dispersed in cholesteric liquid crystals (CLC), and have shown that these can be controlled by tuning the ratio between the particle size $R$ (typically its radius) and the cholesteric pitch $p$. For instance, if strong perpendicular (homeotropic) anchoring is set on the surface of a spherical colloid, one observes either  a planar Saturn ring (a circular defect line of half-integer charge located around the equator of the particle) if $p >> R$, or a twisted Saturn rings, wrapping around the particle, if $p<R$~\cite{lintuvuori2,lintuvuori3,review,musevic2}. If, on the other hand, the anchoring is tangential to the surface, the defect pattern ranges from the "boojum" (two surface defects of integer charge located on opposite sides) if $R/p$ is small, to its twisted version if $R/p$ is large enough~\cite{review}. A further complication arises when, in place of a solid particle, one considers a deformable liquid droplet. Here the strength of the interface anchoring relative to the elasticity of the LC becomes a crucial parameter to control the droplet shape (in addition to the director pattern near the surface). For instance, one observes either a spherical shape or a nutshell-like structure  if the anchoring strength is respectively weak or strong~\cite{Soft_Matter}.  So  far most of the studies have focused on the equilibrium properties of these inclusions~\cite{review} and much less is known when these systems are subject to external perturbations, such as electric and flow fields. 

Liquid crystal have long been known as systems that exhibit a rich dynamical response  such as shear banding~\cite{olmsted,olmsted2,orlandini} and molecular tumbling~\cite{deGennes,tsuji,tsuji2} when subject to  shear flow~\cite{deGennes}. This is due to the complex coupling between hydrodynamics and director field (known as backflow), a feature that is even more relevant in a CLC where the inherent three-dimensional twisted arrangement of the director field can give rise to striking effects, such as a significant increase in viscosity when subject to a shear flow in the direction of its helix (\emph{permeation})~\cite{helfrich,hongla,marenduzzo}. Recent simulations  on colloidal dispersions in CLC report, for instance, a violation of the Stokes law when a single particle is dragged parallel to the cholesteric helix ~\cite{lintuvuori2}, or an induced rotation, either continuous or stepwise, when two particles, forming a dimer, are pulled through the cholesteric phase with a constant force along the helical axis~\cite{juho_dimer}. 

Here we present a preliminary study on the effect that an imposed shear flow has on the rheology of an inverted cholesteric emulsion, described as a single isotropic (liquid crystal) droplet (in which molecules are randomly oriented) surrounded by a cholesteric liquid crystal. The study has been carried out by using a lattice Boltzmann approach~\cite{succi}, a method which solves numerically the Beris-Edwards equations of cholesteric hydrodynamics~\cite{Beris-Edwards} coexisting with an isotropic phase and already tested for an inverted nematic emulsion in the presence of an electric field~\cite{sulaiman} or a shear flow~\cite{Soft_Matter}. The main strength of the algorithm is that backflow effects are automatically included and the dynamics of topological defects can be easily tracked. By varying the shear rate and the ratio between the elastic energy scales of the cholesteric in the bulk and at the droplet interface, we show that the presence of a liquid droplet strongly reduces the secondary flow usually observed in a pure cholesteric phase. Such effect has been found more pronounced when strong perpendicular anchoring is set on the droplet interface, and is also confirmed when the chirality degree of the cholesteric is increased, even though in this case the system may temporarily enter the blue phase for high shear rates~\cite{mermin}.

The paper is organized as follows. In Section 2 we describe the numerical model of a 2D droplet of isotropic fluid suspended in a CLC, in particular its equilibrium phase behavior, encoded by a Landau-de Gennes free energy, and its hydrodynamics, captured by the Beris-Edwards equations of motion for CLCs coupled to Cahn-Hilliard dynamics. In Section 3 we first discuss the equilibrium properties of a cholesteric sample and then those of a single droplet of Newtonian fluid dispersed into it, both in the absence of anchoring and with homeotropic anchoring.  We next present the main results of the paper, namely the effect that a symmetric shear flow has on the droplet-CLC system when different values of interface anchoring and chirality are considered. Finally the last section is dedicated to discuss the results and to conclusions.
 
\section{Model and methods}

We consider an isolated  droplet of Newtonian isotropic fluid dispersed in a medium of cholesteric liquid crystal.  This setup is usually referred as {\it inverted} cholesteric emulsion~\cite{poulin,weitz_PRE,Soft_Matter} to be distinguished from a {\it direct} emulsion, where a liquid crystal droplet is immersed in an isotropic fluid. The hydrodynamics of such system can be described by using an extended version of the Beris-Edwards theory~\cite{Beris-Edwards} for chiral fluids, already adopted in previous works on liquid crystals~\cite{cates_soft,tiribocchi_soft,tiribocchi_prl,tiribocchi_soft2}. Here we briefly recap the model. 

The equilibrium properties are captured by a Landau-de Gennes~\cite{deGennes} free-energy functional ${\cal F}=\int_V{fdV}+\int_S f_WdS$, where the free-energy density $f$ is given by the sum of the following terms
\begin{equation}\label{free_energy}
f= f_{bf} (\phi) + f_{lc} (\phi, Q) + f_{int} (\phi, Q),
\end{equation}
while the second contribution $f_W(Q)$ is added in the presence of a bounding surface. Here $\phi({\bf r},t)$ is a scalar order parameter related to the concentration of the cholesteric phase relative to the isotropic phase, while ${\bf Q}({\bf r},t)$ is a tensor order parameter that, within the Beris-Edwards theory~\cite{Beris-Edwards}, describes the cholesteric phase. It is defined as $Q_{\alpha\beta}=q(\hat{n}_{\alpha}\hat{n}_{\beta}-1/3\delta_{\alpha\beta})$, where $\hat{n}({\bf r},t)$ is the director field\footnote{Greek subscripts denote Cartesian coordinates.} describing the local orientation of the molecules (in the uniaxial approximation $\textbf{n}=-\textbf{n}$), and $q$ is the local degree of order, proportional to the largest eigenvalue of ${\bf Q}$ ($0\le q\le 2/3$). The first term of Eq.~\ref{free_energy} describes the bulk properties of the mixture and is given by 
\begin{equation}\label{free_binary}
f_{bf}(\phi)=\frac{a}{4}\phi^2(\phi-\phi_0)^2+\frac{K_{bf}}{2}|\nabla\phi|^2,
\end{equation}
where $a$ and $K_{bf}$ are two positive phenomenological constants controlling the interface width $\Delta$ of the droplet, which, for a binary fluid without liquid crystal, goes as $\Delta\sim\sqrt{K_{bf}/a}$. Eq.~\ref{free_binary} is borrowed from binary fluid mixtures and enables the formation of two phases: The isotropic one (inside the droplet where $\phi\simeq 0$) and the cholesteric one (outside the droplet where $\phi\simeq\phi_0$), separated by an interface whose energetic cost is gauged by the gradient term. The second term is the cholesteric liquid crystal free-energy density given by
\begin{eqnarray}\label{free_bulk_chol}
f_{lc}(\phi, Q)&=&A_0\left[\frac{1}{2}\left(1-\frac{\zeta (\phi)}{3}\right)Q_{\alpha\beta}^2-\frac{\zeta (\phi)}{3}Q_{\alpha\beta}Q_{\beta\gamma}Q_{\gamma\alpha}+\frac{\zeta (\phi)}{4}(Q_{\alpha\beta}^2)^2\right]\nonumber\\
&&+\frac{K}{2}(\partial_{\beta}Q_{\alpha\beta})^2+\frac{K}{2}(\epsilon_{\alpha\delta\gamma}\partial_{\delta}Q_{\gamma\beta}+2q_0Q_{\alpha\beta})^2.
\end{eqnarray}
The terms multiplied by the positive constant $A_0$ stem from a truncated polynomial expansion up to the fourth order in ${\bf Q}$~\cite{deGennes}, and describe the bulk properties of an uniaxial nematic liquid crystal with an isotropic-to-nematic transition at $\zeta(\phi)=\zeta_c=2.7$. Here $\zeta$ plays the role of an effective temperature. For a nematogen without chirality ($q_0=0$), the phase is isotropic if $\zeta(\phi)<\zeta_c=2.7$, otherwise it is cholesteric. By following previous studies~\cite{sulaiman,Soft_Matter,Fadda} we set $\zeta(\phi)=\zeta_0+\zeta_s\phi$, where $\zeta_{0}$ and $\zeta_{s}$ control the boundary of the coexistence region. The remaining terms of Eq.~\ref{free_bulk_chol}, multiplied by the constant $K$, take into account the elastic energy due to the local deformations of the cholesteric arrangement and enter the free energy through first order gradient contributions, except a gradient-free term included to have a positive elastic free energy. The parameter $q_0=2\pi/p_0$ determines the pitch length $p_0$ of the cholesteric and $\epsilon_{\alpha\delta\gamma}$ is the Levi-Civita antisymmetric tensor. Here we consider the ``one elastic constant'' approximation, an approach usually adopted when investigating liquid crystals as it considerably simplifies theoretical calculations~\cite{deGennes}. The energetic cost due to the anchoring of the liquid crystal at the droplet interface is included through
\begin{equation}
f_{int}(\phi,Q)=L(\partial_{\alpha}\phi)Q_{\alpha\beta}(\partial_{\beta}\phi),
\end{equation}
where the constant $L$ controls the strength of the anchoring. If it is negative the liquid crystals is homeotropic (perpendicular) to the interface, whereas if positive the liquid crystal lies tangentially. Finally, if confining walls are included, a further term needs to be considered in the free energy functional. For homeotropic  anchoring, which is the case considered here, one has
\begin{equation}
f_W(Q)=\frac{1}{2}W(Q_{\alpha\beta}-Q^0_{\alpha\beta})^2,
\end{equation}
where $W$ controls the strength of the anchoring at the walls and $Q_{\alpha\beta}^{0}$ sets the preferred configuration of the tensor order parameter at the surface.

The thermodynamic state of our mixture is specified by two dimensionless quantities, the reduced temperature  and the reduced chirality, given by
\begin{eqnarray}
\tau&=&\frac{27(1-\zeta(\phi_0)/3)}{\zeta(\phi_0)},\label{red_temp}\\
\kappa&=&\sqrt{\frac{108Kq_0^2}{A_0\zeta(\phi_0)}}.\label{chiral}
\end{eqnarray}
The former multiplies the quadratic terms of the dimensionless bulk free energy and vanishes at the spinodal point of a nematic, while the latter multiplies the gradient terms and gauges the amount of twist accumulated in the system~\cite{mermin}. Such parameters have been used, for instance, in Refs.~\cite{dupuis,henrich2} for the numerical calculation of the  phase diagram of cholesteric liquid crystals. Note, in particular, that according to Eq.~\ref{chiral} the knowledge of $q_0$ is not a sufficient information (except if $q_0=0$) to correctly assess the value of the chirality (hence the phase of the liquid crystal), as this is also affected by other thermodynamic parameters ($A$, $K$ and $\zeta$).

The dynamics of the system is governed by a set of balance equations, the first of which is the equation of the tensor order parameter ${\bf Q}$ 
\begin{equation}\label{Q_eq}
(\partial_t+{\bf u}\cdot\nabla){\bf Q}-{\bf S}({\bf W},{\bf Q})=\Gamma {\bf H},
\end{equation}
where ${\bf u}$ is the velocity of the fluid and the term on the left hand side is the material derivative, describing the rate of change 
of ${\bf Q}$ advected by the flow. The derivative includes the tensor ${\bf S}({\bf W},{\bf Q})$ since the order parameter can be rotated and stretched by local velocity gradients $W_{\alpha\beta}=\partial_{\beta}u_{\alpha}$, and is given by
\begin{equation}\label{eq_S}
{\bf S}({\bf W},{\bf Q})=(\xi{\bf D}+{\bf \Omega})({\bf Q}+{\bf I}/3)+({\bf Q}+{\bf I}/3)(\xi{\bf D}-{\bf \Omega})-2\xi({\bf Q}+{\bf I}/3)Tr({\bf Q}{\bf W}).
\end{equation}
Here $\textbf{D}=(\textbf{W}+\textbf{W}^{T})/2$ and ${\bf \Omega}=(\textbf{W}-\textbf{W}^{T})/2$ are the symmetric and antisymmetric parts of the velocity gradient tensor, $Tr$ denotes the tensorial trace and  $\textbf{I}$ is the unit matrix. The constant $\xi$ depends upon the molecular details of the liquid crystal, and controls the dynamics of the director field under shear flow. Indeed after imposing a homogeneous shear on a nematic liquid crystal, at steady state the director will align along the flow gradient at an angle $\theta$ fulfilling the relation $\xi\cos(2\theta)=(3q)/(2+q)$. Real solutions are obtained when $\xi\geq 0.6$. Throughout our simulations we have set $\xi=0.6$. Finally, on the right hand side of Eq.~\ref{Q_eq}, $\Gamma$ is the collective rotational diffusion constant 
and ${\bf H}$ is the molecular field given by
\begin{equation}
{\bf H}=-\frac{\delta {\cal F}}{\delta {\bf Q}}+\frac{{\bf I}}{3}Tr\frac{\delta {\cal F}}{\delta {\bf Q}}.\label{H_eq}
\end{equation}
The time evolution of the concentration field $\phi({\bf r},t)$ is governed by a convection-diffusion equation
\begin{equation}
\partial_{t}\phi+ \partial_{\alpha}(\phi u_{\alpha})=\nabla \left (M \nabla\frac{\delta {\cal F}}{\delta \phi}\right),\label{phi_eq}
\end{equation}
where $M$ is the mobility and $\mu=\delta {\cal F}/\delta\phi$ is the chemical potential, while the force balance is ensured by the incompressible Navier-Stokes equation,
\begin{eqnarray}
\nabla\cdot{\bf u}&=&0,\label{cont}\\
\rho(\partial_t+u_{\beta}\partial_{\beta})u_{\alpha}&=&\partial_{\beta}\sigma_{\alpha\beta}^{total},\label{nav_stok}
\end{eqnarray}
where the total stress tensor $\sigma^{total}_{\alpha\beta}$ is the sum of four contributions
\begin{equation}
\sigma^{total}_{\alpha\beta}=-\rho T+\sigma^{visc}_{\alpha\beta}+\sigma^{lc}_{\alpha\beta}+\sigma^{s}_{\alpha\beta}.
\end{equation}
The first term $\rho T$ is the ideal background pressure and $T$ is the temperature. The second one is the viscous stress tensor
\begin{equation}
\sigma^{visc}_{\alpha\beta}=\eta(\partial_{\alpha}u_{\beta}+\partial_{\beta}u_{\alpha}),
\end{equation}
where $\eta$ is the isotropic shear viscosity. The third one is the elastic stress due to the liquid crystalline order
\begin{eqnarray}\label{lc_stress}
  \sigma^{lc}_{\alpha\beta}&=&\frac{K}{2}(\nabla\textbf{Q})^2\delta_{\alpha\beta}-\xi H_{\alpha\gamma}\left ( Q_{\gamma\beta}+\frac{1}{3}\delta_{\gamma\beta} \right )-\xi\left ( Q_{\alpha\gamma}+\frac{1}{3}\delta_{\alpha\gamma} \right )H_{\gamma\beta}\nonumber\\
&&+2\xi\left ( Q_{\alpha\beta}-\frac{1}{3}\delta_{\alpha\beta} \right )Q_{\gamma\mu}H_{\gamma\mu}+Q_{\alpha\nu}H_{\nu\beta}-H_{\alpha\nu}Q_{\nu\beta},
\end{eqnarray}
and the last term is the interfacial stress between the isotropic and liquid crystal phase
\begin{equation}\label{int_stress}
\sigma_{\alpha\beta}^{s}=-\left(\frac{\delta{\cal F}}{\delta\phi}\phi-{\cal F}\right)\delta_{\alpha\beta}-\frac{\delta{\cal F}}{\delta(\partial_{\beta}\phi)}\partial_{\alpha}\phi-\frac{\delta{\cal F}}{\delta(\partial_{\beta}Q_{\gamma\mu})}\partial_{\alpha}Q_{\gamma\mu}.
\end{equation}
The total isotropic pressure of the system is the sum of the ideal background pressure $\rho T$ plus the isotropic term of Eq.~\ref{lc_stress} (constant in our simulations, except nearby the defects and close to the droplet interface where it augments) and of Eq.~\ref{int_stress}.
Both $\sigma^{lc}_{\alpha\beta}$ and $\sigma_{\alpha\beta}^{s}$ take into account the non-Newtonian fluid effects. In fact, if ${\bf Q}=0$, they reduce, respectively, to the scalar pressure and to the interfacial stress of a binary Newtonian fluid mixture.

Equations~(\ref{Q_eq})~(\ref{phi_eq})~(\ref{cont})~(\ref{nav_stok}) are solved numerically by means of a hybrid lattice Boltzmann method~\cite{henrich,Fadda}, which uses a combination of a standard lattice Boltzmann approach to solve the Navier-Stokes and the continuity equation, and a finite-difference scheme to solve Eqs.~(\ref{Q_eq}) and (\ref{phi_eq}).

All simulations are performed on a quasi two-dimensional rectangular box of size $L_{x}=1 \times L_{y}=300 \times L_{z}=100$ in which the cholesteric liquid crystal and the droplet are embedded. We choose a longer size along the $y$-direction in order to minimize the interference between periodic images of the droplet moving with the flow when a shear flow is applied. The droplet, in particular, is modeled as a circular isotropic (liquid-crystalline) region with radius $R=32$ lattice sites, initially placed at the centre of the sample and surrounded by the liquid crystal phase. This quasi-2d setup is chosen to allow an out-of-plane component of the vector fields along the $x$-direction that accommodates the three-dimensional twisted arrangement of the cholesteric, and, in the presence of a shear flow, captures the secondary flow appearing perpendicular to the direction of the shearing, along the $x$-direction. 
The entire system is periodic along the $y$-direction and is sandwiched between two flat walls parallel to the $xy$-plane and lying at $z=0$ and $z=L_z$. The walls can be either at rest or moving along the $y$-direction with velocity $-u_w$ and $u_w$ (with $u_w>0)$ at $z=0$ and $z=L_z$, respectively. There we impose neutral wetting conditions (i.e. ${\bf a}\cdot\nabla\phi=0$, where ${\bf a}$ is a unit vector perpendicular to the wall) for the concentration, no-slip conditions (the fluid moves with the same velocity of the walls) for the fluid velocity and homeotropic conditions (i.e. the director is perpendicular to the walls) for the liquid crystal order parameter.

Initial conditions of $\phi$ and ${\bf Q}$ in the bulk are as follows. They are both set to zero inside the droplet, while outside $\phi$ is fixed to a constant value ($\phi_{0}\simeq 2$) and the components of $\textbf{Q}$ are given by
\begin{eqnarray}
Q_{xx}&=& (c_0-c_1/2)\cos(2q_0y)  + c_1/2,\\
Q_{yy}&=& -2c_1,\\
Q_{xz}&=& -(c_0-c_1/2)\sin(2q_0y),\\
Q_{xy}&=&Q_{yz}=0,
\end{eqnarray}
where $c_0=0.546$, $c_1=0.272$, to accommodate a cholesteric liquid crystal with helical axis parallel to the $y$-direction. 
Such components stem from assuming that the ground state of the director field is given by $\hat{\bf n}(y)=\cos(q_0y){\bf e}_x+\sin(q_0y){\bf e}_z$, where ${\bf e}_{\alpha}$ is a unit vector oriented along the Cartesian axis. The parameter $q_0 = 2\pi/p_0$ controls the number $N$ of  $\pi$ twists that the liquid crystal displays in a cell of length $L_y$. Indeed one can define $p_0$ in terms of the linear size $L_y$, as $p_0=2L_y/N$, to compare the pitch length with the cell size. 

Starting from these initial conditions, the system is first let to relax to its equilibrium state and afterwards both walls are sheared along opposite directions at constant speed. Typical parameters used in our simulations are $a=7 \times 10^{-2}$, $K_{bf}=0.14$, $M=5 \times 10^{-2}$, $\Gamma=1$, $A_{0}=0.1$, $K=0.1$, $\eta=0.38$, $L=-0.04$ and $W=0.4$. Finally one has to choose the values of $\tau$ and $\kappa$ in order to have the cholesteric phase outside the droplet~\cite{dupuis,henrich2}. We set $\tau=0$ (i.e. $\zeta(\phi_0)=3$) and $q_0$ (which controls the chirality $\kappa$, see Eq.~\ref{chiral}) equal to $\pi/150$ or $\pi/75$, in order to have $N=2$ or $N=4$ on a lattice of linear size $L_y=300$. If, for instance, $q_0=\pi/75$, one has $\kappa~\simeq 0.25$, well inside the cholesteric phase according to the phase diagram of Ref.~\cite{dupuis,henrich2}.

In order to map such numerical values to typical physical ones, we consider a droplet of size $1-10\mu$$m$ immersed in a cholesteric liquid crystal of elastic constant roughly $10$pN and viscosity of $1$ Poise. This corresponds to a lattice space $\Delta x=10^{-7}$m and the time step is $\Delta t=10^{-8}$s (they are both equal to one in our simulations). Finally the anchoring coefficient $L$ correspond to values $10^{-6}$J$m^{-2}$~\cite{sulaiman,poon}.

\section{Results}

We investigate the physics of two different systems, one in which the sole cholesteric liquid crystal is sandwiched between flat walls and the other one in which an isotropic droplet is dispersed inside it. We initially study their equilibrium properties, in particular when homeotropic anchoring of the director is set both at the walls and at the droplet interface, and for two different values of the cholesteric pitch. Afterwards the dynamic response under different shear rates (only in the flow aligning regime) and interface anchoring is studied.

\subsection{Equilibrium properties}

\begin{figure*}[htbp]
\centerline{\includegraphics[width=1.0\textwidth]{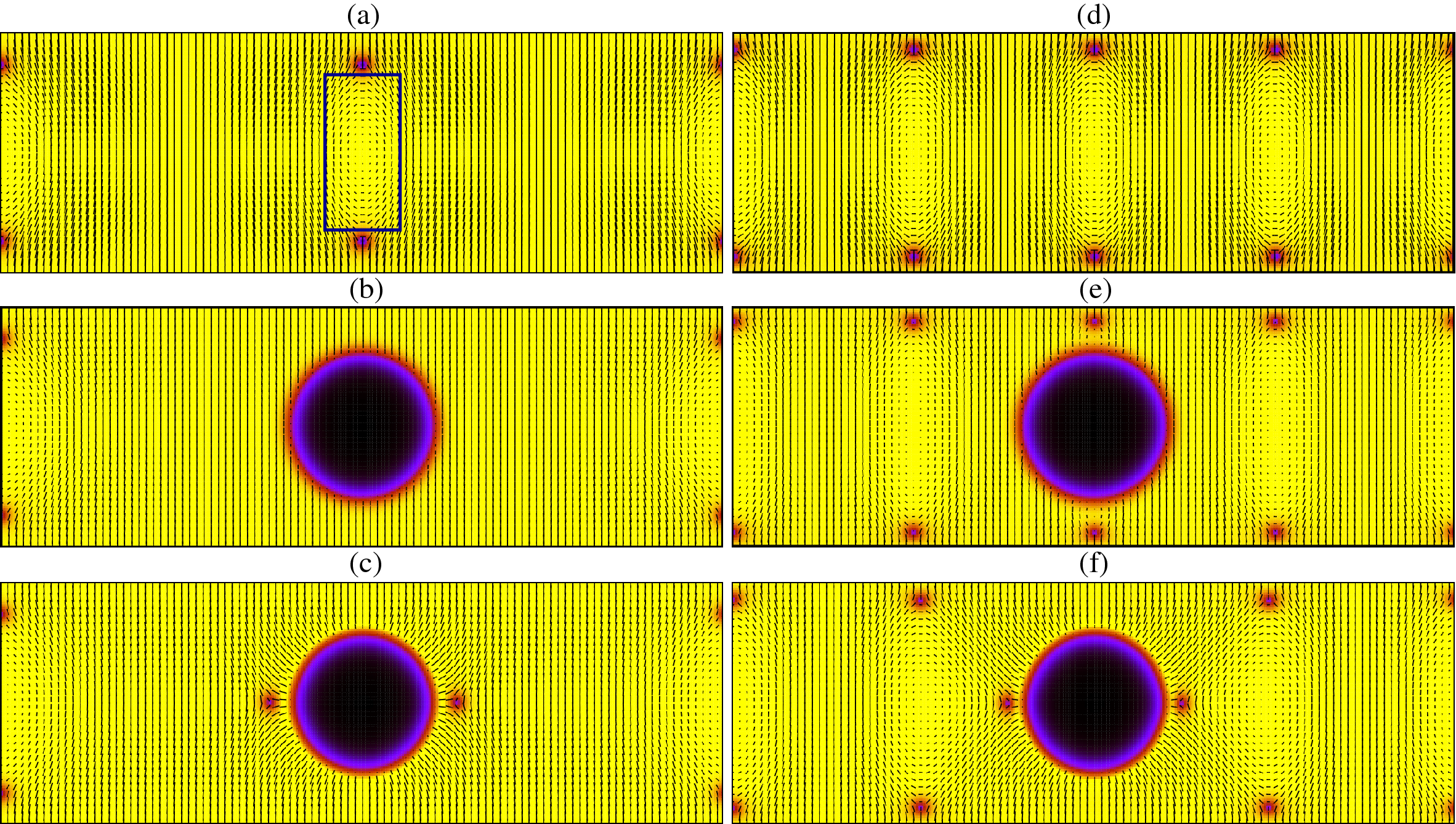}}
\caption{(a)-(d) Equilibrium configurations of the director field of a cholesteric liquid crystal in a cell of size $L_y=300$ and $L_z=100$ lattice units, for $N=2$ (a) and $N=4$ (d) $\pi$-twists of the helical axis. In both cases homeotropic anchoring is set at the walls. A $\lambda^{+1}$-charge region is highlighted a by continuous line in (a). (b)-(c)-(e)-(f) Equilibrium configurations of an isotropic droplet surrounded by a cholesteric liquid crystal. The radius of the droplet is $R=32$ lattice units and the cholesteric pitch has $N=2$ (b,c) and $N=4$ (e,f) $\pi$-twists. While in (b) and (e) the interface anchoring is $L=0$, in (c) and (f) $L=-0.04$ to enforce homeotropic anchoring. The color map represents the largest eigenvalue of the ${\bf Q}$ tensor and ranges from $0$ (black) to $\simeq 0.33$ (yellow). $tau$ disclinations are highlighted by blue-red spots as the scalar order parameter drops down, while it is approximately constant in rest of the cholesteric.}
\label{fig1}
\end{figure*}

We first consider a CLC initialized in a quasi two-dimensional rectangular box of size $L_{y}=300 \times L_{z}=100$ with walls at rest. In Fig.~\ref{fig1}(a) and (d) we show the equilibrium configurations of the director field (obtained after $t~\simeq 10^5$ timesteps, when the free-energy is at its minimum value) of the cholesteric phase for $N=2$ (a) and $N=4$ (d) and with strong homeotropic anchoring on both walls. In both cases the cholesteric axis lies along the $y$-direction. One can clearly distinguish regions where the director displays a local nematic-like order (with orientation almost entirely along the $z$-direction, forced by the anchoring at the walls) from regions where it flips in the third direction (i.e. along the $x$-axis). Due to the interplay between the  orientation of the cholesteric phase near the walls and the imposed perpendicular anchoring, topological defects emerge. Although in cholesterics four different types of defects may occur~\cite{Fadda,kleman,kleman2}, here we observe  $\lambda^{\pm m}$ and $\tau^{\pm m}$ disclinations defects ($m$ is the topological charge, an integer or semi-integer number). While in $\lambda$ defects the director field is continuous (it is the local cholesteric pitch axis, winding around the defect, that is discontinuous) and flips into the third dimension at the $\lambda$ line, $\tau$ defects are true disclinations as the director drops down at the defect core. Note that in  Fig.~\ref{fig1}(a) and (d) those pinned at the walls are $\tau$ disclinations (of charge $-1/2$) which sustain a $\lambda^{+1}$-charge region, where the director field flips into the third dimension. Changing the chirality also affects their position; indeed, when $N=4$, $\tau$ disclinations appear closer to the boundaries and $\lambda$-charge regions are longer and narrower than those observed when $N=2$. It important to stress  that, while it is easy to  track numerically the location of $\tau$ disclination defects (for instance by computing the local orientational order at each point), it is more difficult to resolve the correct position of $\lambda$ defects, as the radius of their core is usually comparable with the helix pitch $p$, and only a limited number of lattice points can be used to calculate its extension. This is why we prefer indicating regions containing $\lambda$ defects (either of charge $\pm 1$ or couple of charge $\pm 1/2$ ) also as $\lambda$-charge regions. 

The presence of an isotropic droplet in the centre of the cell substantially modifies the equilibrium configuration of the cholesteric. In the absence (or for very weak values) of interface anchoring and with $N=2$ (Fig.~\ref{fig1}(b)), the droplet removes the $\lambda$-charge region and the two $\tau$ disclinations in the centre of the box leaving the rest of the cholesteric almost unaltered. If, on the other hand $N=4$, the two $\tau$ disclinations in the centre of the box survive (as they are closer to the walls), and the $\lambda$-charge region is simply covered by the droplet (Fig.~\ref{fig1}(e)). If strong homeotropic anchoring is imposed at the droplet interface (Fig.~\ref{fig1}(c)-(f)), two fully in-plane topological defects of charge $-1/2$ emerge on opposite sides of the droplet and located along the equatorial line. Such strong anchoring importantly alters the liquid crystal orientation in the surroundings of the droplet, favoring the formation of a pronounced splay deformation.

A suitable quantity to measure the strength of the interface anchoring relative to the bulk elastic deformation is the dimensionless number $\Lambda={\cal F}_{int}R/\Sigma K$, where ${\cal F}_{int}=\int_V f_{int}dV$ and $\Sigma$ is the perimeter of the droplet. For a droplet of size $1-10\mu$m, $\Lambda$ usually ranges from $10^{-2}$ to $10^3$. If $\Lambda\ll 1$ the anchoring is weak, and no large deformations are observed in the surroundings of the droplet (as in Fig.~\ref{fig1}(b)-(e), where $\Lambda\simeq 0$), whereas if $\Lambda\sim 1$ the anchoring is considered strong and large distortions of director with topological defects are found (see Fig.~\ref{fig1}(c)-(f), where $\Lambda\simeq 2.88$).

\subsection{Cholesteric liquid crystal under shear}

\begin{figure*}[htbp]
\centerline{\includegraphics[width=1.0\textwidth]{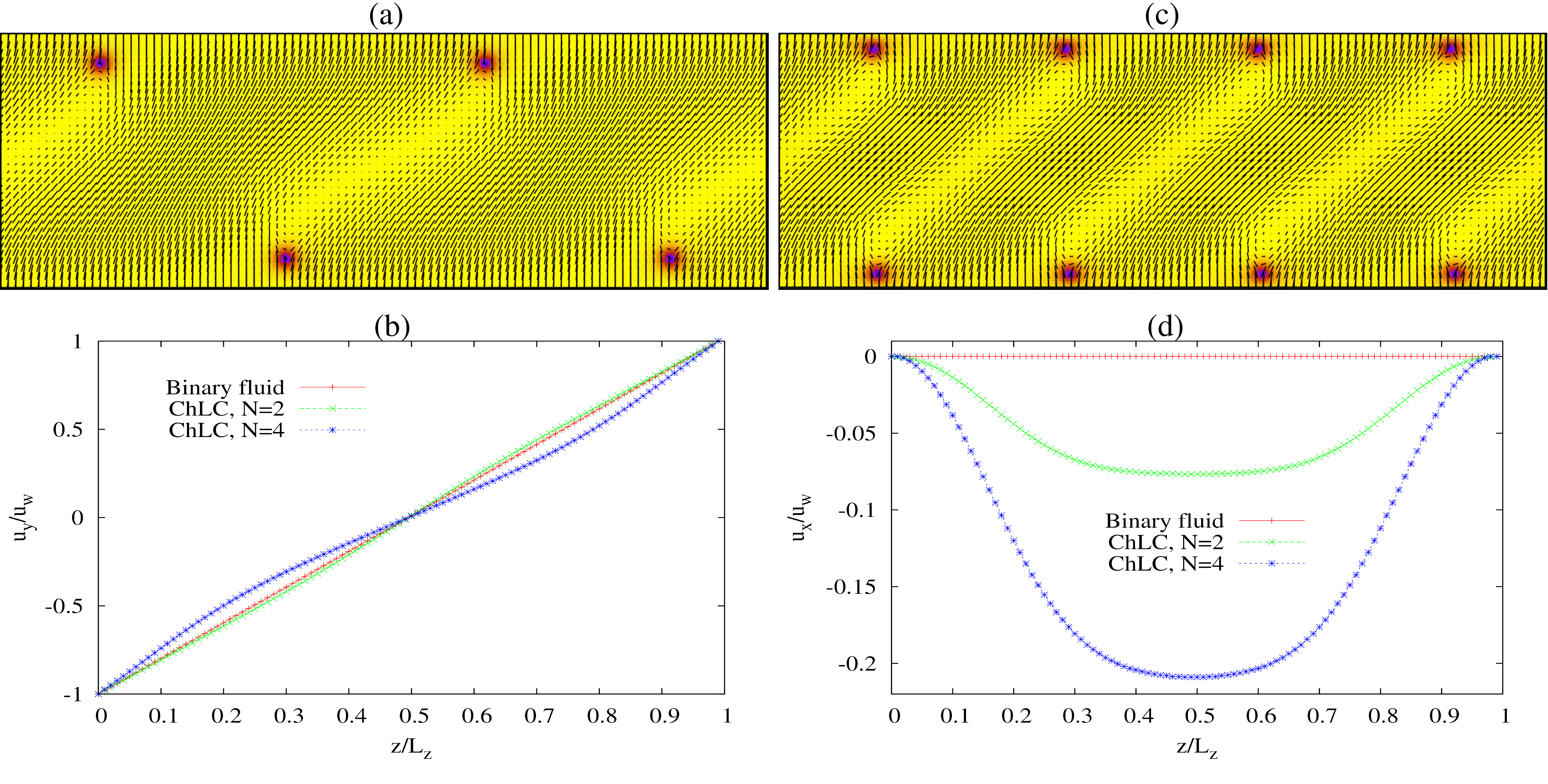}}
\caption{(a)-(c) Steady-state configurations of the director field for $N=2$ (a) and $N=4$ (c) observed when a relatively weak symmetric shear flow ($\dot{\gamma}=6\times 10^{-5}$) is applied. (b)-(d) Profiles of the velocity field $u_y$ (primary flow) and $u_x$ (secondary flow) taken at $y=L_y/2$ for a binary fluid (${\bf Q}=0$) (red) and a cholesteric liquid crystal with $N=2$ (green) and $N=4$ (blue).}
\label{fig2}
\end{figure*}

By starting from the equilibrated configurations previously described, we now study the rheology of such systems when a symmetric shear flow is imposed. We move top and bottom walls along the $y$-axis with velocity $u_w$ (top) and $-u_w$ (bottom). This sets a shear rate $\dot{\gamma} = 2 u_w/L_z$ (measured in $\Delta t^{-1}$ in simulation units) and a shear flow along the $y$-direction. We first consider the homogeneous case in which no droplets are present. In Fig.~\ref{fig2}(a)-(c) we show the steady state configurations of the director field for $N=2$ and $N=4$ and for a relatively weak shear rate ($\dot{\gamma}=6\times 10^{-5}$, i.e. $u_w=0.003$). With respect to the equilibrium state, the cholesteric layers are tilted along the direction imposed by the shear flow, while $\tau$ disclinations, still pinned at the walls, sustain $S$-shaped $\lambda$-charge regions. The primary flow $u_y$ (measured at $y=L_y/2$) is approximately Newtonian for $N=2$ while it gets weakly sigmoidal  $N=4$ (Fig.~\ref{fig2}b), an effect due to the larger resistance to overcome elastic deformations encountered by the flow when higher values of $q_0$ are considered. Note that, as reported in Ref.~\cite{marenduzzo}, a more flattened velocity profile (in which $u_y$ is zero almost everywhere along the $z$-direction) could be achieved if larger values of $L_z$ and shorter values of $L_y$ are considered. As expected, unlike the case of a binary fluid mixture without liquid crystal (i.e. when ${\bf Q}=0$), a secondary flow $u_x$ emerges due to the action of the director field on the velocity field along the $x$-direction (Fig.~\ref{fig2}(d)), stronger in the center of the sample (where it attains its larger value) but smaller than the primary flow, and zero near the walls. Its negative sign is due to the handedness of the initial configuration of the liquid crystal. Note that the presence of a secondary flow is a signature of the permeation, i.e. a significant increase in the viscosity (calculated as $\eta_{yz}=\sigma_{yz}^{total}/\partial_zu_y$~\cite{marenduzzo}) of the cholesteric liquid crystal observed when it is subject to a flow in the direction of its helix (i.e. perpendicular to the director field)~\cite{deGennes,helfrich,lubensky,marenduzzo}. Our simulations show, for instance, that, if $N=4$, $\eta_{yz}$ attains a maximum at $\sim 0.65$ for $\dot{\gamma}=6\times 10^{-5}$ (in the centre of the sample) and at $\sim 1.4$ for $\dot{\gamma}=8\times 10^{-5}$. Such values are higher than the isotropic viscosity $\eta=0.38$.

\begin{figure*}[htbp]
\centerline{\includegraphics[width=1.0\textwidth]{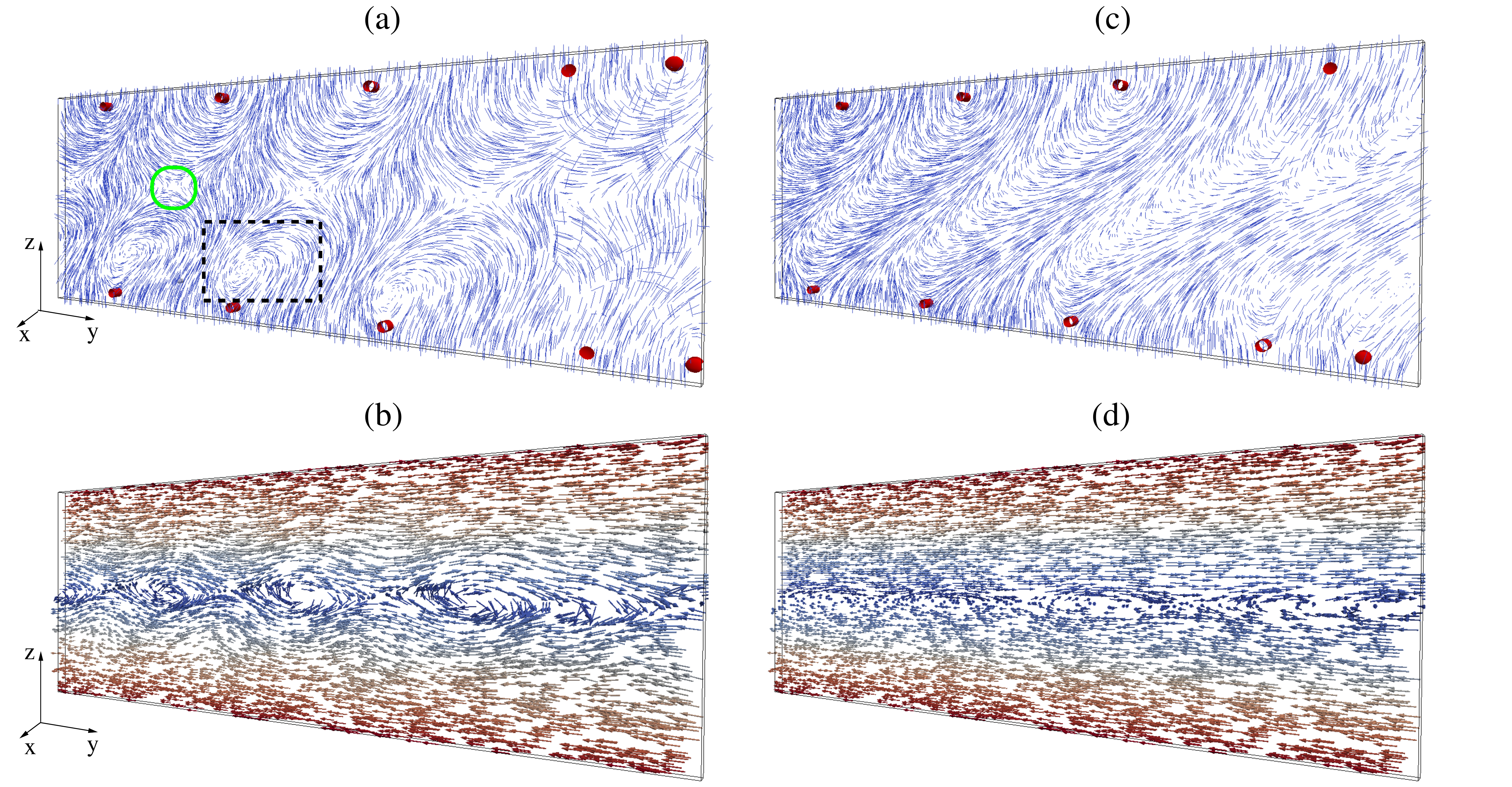}}
\caption{(a)-(c) Dynamic configurations of the director field observed when a symmetric shear rate $\dot{\gamma}=2\times 10^{-4}$ is applied to a cholesteric sample with $N=4$ in a 3D system of size $L_x=5$, $L_y=300$ and $L_z=100$. Here $t_a<t_c$, where $t_a$ and $t_c$ indicate the simulation times of (a) and (c), respectively. Red columns indicate regions where the scalar order parameter drops down and where $\tau^{1/2}$ disclinations form. The green circle indicate a $\lambda^{-1}$-charge region whereas the black dotted square a double twist cylinder. In (b) and (d) the velocity fields corresponding to panels (a) and (c), respectively, are shown. The color of the arrows ranges from $u=0.01$ (at the walls) to $u\simeq 10^{-5}$ in the middle of the cell.}
\label{fig3}
\end{figure*}

A well-established result of the rheology of liquid crystals is that if the shear-rate is increased, the temperature at which the cholesteric-nematic transition occurs actually diminishes~\cite{olmsted,marenduzzo}. While we confirm this result, we also find that the transition depends upon the pitch length, namely by $N$. In order to better appreciate it, in Fig.~\ref{fig3} we show two dynamic configurations of the director field and of the corresponding velocity field for $N=4$ and $\dot{\gamma}=2\times 10^{-4}$ ($u_w=0.01$), in a full three dimensional box of size $L_x=5$, $L_y=300$, $L_z=100$. While if $N=2$ (and for the same shear rate) the cholesteric is already in the nematic state (not shown), if $N=4$ the system exhibits a more complex dynamics, in which double-twist cylinders (one of them is highlighted with a black dotted square in Fig.~\ref{fig3}a) periodically form and get destroyed by the fluid flow. Afterwards the cholesteric arrangement is temporarily restored (Fig.~\ref{fig3}c) and disrupted again. The presence of double twist cylinders indicate that, within a range of values of shear-rate, the cholesteric-nematic transition is replaced by the cholesteric-blue phase one for high values of $N$ (namely for a higher chirality). Interestingly, during the blue-phase-like regime topological defects at the walls acquire a positive charge (they are now actually $\tau^{1/2}$ disclinations, indicated by red cylinders in Fig.~\ref{fig3}a) while the topological charge of the $\lambda$ region (highlighted by a green circle in Fig.~\ref{fig3}a) switches to $-1$. The velocity field shows large vortices in the surroundings of such defects (Fig.~\ref{fig3}b) while far from them it has the typical structure of sheared systems (such as in Fig.~\ref{fig3}d), i.e. a bidirectional flow, strong near the walls and much weaker in the middle of the cell.

\subsection{Isotropic droplet in cholesteric phase under shear}

\noindent We now describe the rheological properties of a system in which an isotropic droplet is included in the cholesteric phase. We consider low/moderate values of shear rate, comparable to those used for the cholesteric cell without droplet. This also ensures that the system remains firmly in the cholesteric phase and finite size effects due to periodic boundary conditions and artificial interactions between droplet mirror images  should also be minimized. Two dimensionless numbers useful to quantify the effects of the shear flow on the droplet are the Capillary number $Ca$ and the deformation parameter $D$. The former is defined as $Ca=\frac{R\dot{\gamma}\eta\Sigma}{{\cal F}_{K_{bf}}}$ (where $F_{K_{bf}}=\int_V dV (K_{bf}/2)|\nabla\phi|^2$) and measures the strength of viscous forces relative to the surface tension mediated by the constant $K_{bf}$, while the latter is defined as $D=\frac{a-b}{a+b}$~\cite{taylor} (where $a$ and $b$ are the major and the minor axis of an elliptical droplet) and quantifies the droplet deformation under shear. If low or moderate shear flows are considered, $Ca$ is expected to be $\sim 0.01$ while $D$ between $0$ (no deformation) and $0.1$ (weak/moderate deformation).
\begin{figure*}[htbp]
\centerline{\includegraphics[width=1.0\textwidth]{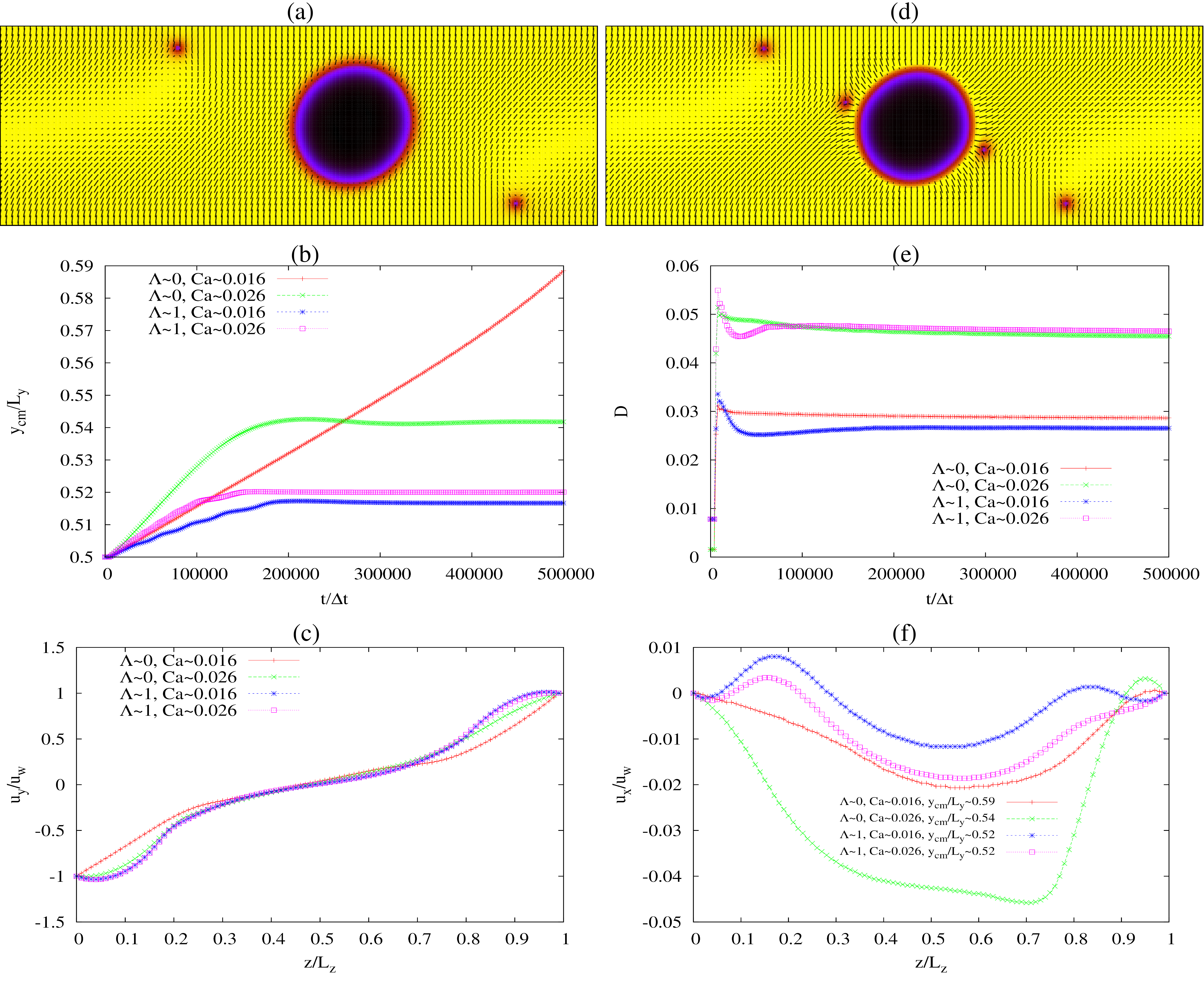}}
\caption{(a)-(d) Two late-time configurations of the director field and of the largest eigenvalue of ${\bf Q}$ observed when a symmetric shear rate $\dot{\gamma}=6\times 10^{-5}$ ($u_w=0.003$, $Ca\sim 0.016$) is applied to a cholesteric sample with $N=2$ for $\Lambda\sim 0$ (a) and 
$\Lambda\sim 1$ (d). Red spots indicate the position of topological defects. Those near the walls are $\tau^{-1/2}$ disclinations while the ones close to the droplet interface are in-plane defects of topological charge $-1/2$. In (b) and (d) the $y$-coordinate of the center of mass of the droplet and the deformation parameter $D$ are shown for different values of $\Lambda$ and shear rates $\dot{\gamma}=6\times 10^{-5}$ ($Ca\sim 0.016$) and $\dot{\gamma}=10^{-4}$ ($Ca\sim 0.026$). Finally in (c) and (f) the primary flow ($u_y$) and the secondary flow ($u_x$), measured at $y=L_y/2$, are shown for the same values of $\Lambda$ and shear rate at late times.} 
\label{fig4}
\end{figure*}

We first consider the case with $N=2$. In Fig.~\ref{fig4}(a)-(d) we show two configurations of the director field for $\Lambda\sim 0$ (a) and $\Lambda\sim 1$ (d) under shear flow with $\dot{\gamma}=6\times 10^{-5}$ ($u_w=0.003$). The starting equilibrium states are those shown in Fig~\ref{fig1}(b) and (c), respectively. While if $\Lambda\sim 0$ the director field in the bulk of the cell is almost everywhere tilted along the direction imposed by the shear flow (except where it flips in the $x$-direction), with $\Lambda\sim 1$ this occurs only far from the droplet as the two defects near the droplet interface foster the formation of large splay distortions. Note that such defects, although still on opposite sides of the droplet as in Fig.~\ref{fig1}(c), have been dragged by the fluid flow around the interface in a clockwise direction. While in this case the droplet has slightly moved from its initial position rightwards and then has got stuck in the steady state with $y_{cm}$ not changing, if $\Lambda\sim 0$ it has acquired a persistent unidirectional motion rightwards for the lower shear rate (see Fig.~\ref{fig4}(b), where the $y$-position of the center of mass of the droplet is plotted). When increasing the shear rate, $y_{cm}$ reaches a steady value. This unidirectional motion occurs as the center of mass of the droplet shifts slightly upwards with respect to its initial position $z=L_z/2$, enough to move the droplet where the flow (in the upper part of the lattice) can drag it rightwards. The symmetric case, in which the droplet is dragged leftwards, would occur if its barycenter were shifted downwards. 
The deformation $D$ (see Fig.~\ref{fig4}(e)) is only weakly affected by the value of $\Lambda$, while the droplet becomes more elliptical when increasing the shear rate. More important, instead, is the effect that the inclusion of a droplet has on the fluid flow profile. Unlike the case of the pure cholesteric cell, here the primary flow ($u_y$) flattens at zero at the center of the sample with an effective shear rate smaller than the imposed one, and the secondary flow ($u_x$) is almost one order of magnitude lower than that of the cholesteric cell. However, while it increases by augmenting $Ca$ if $\Lambda\sim 0$, it is strongly reduced if $\Lambda\sim 1$. The former occurs as, despite an increase in the shear rate, the liquid crystal is still firmly in the cholesteric phase where the secondary flow is expected to be relevant. The latter instead is due to the strong anchoring of the director at the droplet interface and at the walls, a constraint that forces the director to remain fully in the two-dimensional $yz$-plane.

\begin{figure*}[htbp]
\centerline{\includegraphics[width=1.0\textwidth]{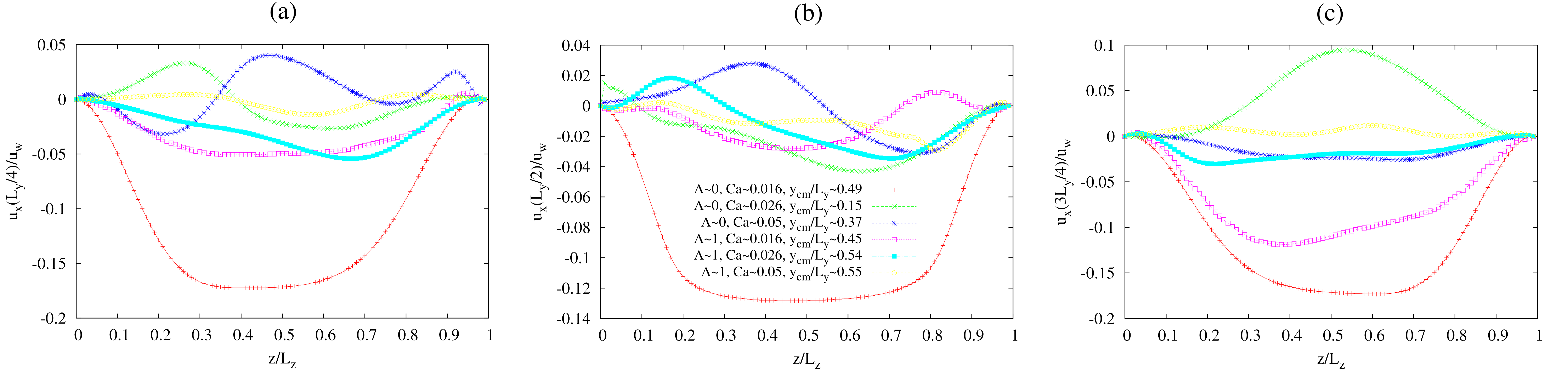}}
\caption{Secondary flow ($u_x$) measured at $y=L_y/4$ (a), $L_y/2$ (b), and $3/4L_y$ (c) for two values of interface anchoring strength 
($\Lambda\sim 0$ and $\sim 1$) and for $Ca=0.016$, $0.026$ and $0.05$ when $N=4$. While $|u_x|$ is high for $\Lambda\sim 0$ and for 
low values of $Ca$, it diminishes by increasing shear rate and it is almost absent for strong interface anchoring.} 
\label{fig5}
\end{figure*}

We now briefly turn to discuss the physics observed when $N=4$. While for low/moderate shear rate the droplet undergoes weak deformations, overall analogous to those shown for $N=2$, for higher values it is generally more difficult to achieve a stationary state as the large number of topological defects favors the formation of intense flows which significantly affect the liquid crystal orientation (through the backflow effect). Despite this limitation, it is interesting to track the secondary flow by varying the shear rate (even out of the cholesteric phase) and the interface anchoring strength in order to assess how the rheological response is affected by the chirality. In Fig.~\ref{fig5} we show the secondary flow profiles taken at three different positions (namely at $y=L_y/4, L_y/2$, and $3L_y/4$) at late times, to give a uniform view of the behavior even in regions where the droplet is absent. As long as $\Lambda$ and $Ca$ are relatively weak ($\Lambda\sim 0$, $Ca\sim 0.016$), $u_x$ attains values comparable with those observed in a sample of cholesteric and almost one order of magnitude higher than those obtained for $N=2$, suggesting that increasing the chirality essentially favors a non-Newtonian response of the fluid. In these cases the droplet achieves a stationary value for $y_{cm}$ (without any appreciable motion) with a deformation overall comparable with that seen for $N=2$. 
Interestingly though, increasing both shear rate and interface anchoring strength leads to a drastic reduction of the secondary flow as, on one hand, the cholesteric order is gradually destroyed in favor of a nematic-like one (where there is no secondary flow) and, on the other hand, the strong interface anchoring impedes the director field to acquire an out-of-plane component (as seen when $N=2$).

\section{Conclusions}

To summarize, we have studied, by numerical simulations, the rheological response of an inverted cholesteric droplet sandwiched between two planar walls under a symmetric shear flow. We have shown that the dynamics is affected by the shear rate, the strength of the interface anchoring and the chirality of the cholesteric. In particular the presence of a relatively intense secondary flow (emerging out of the plane of the cholesteric) suggests that the system is essentially non-Newtonian, an effect generally weakened if an isotropic droplet is included and further reduced if strong interface anchoring is imposed. 

If the droplet is absent, a moderate shear flow drives the sample towards a steady state in which the director field is almost everywhere tilted along the direction imposed by the shear itself, except in regions where $S$-like $\lambda$-charge region are sustained by $\tau$ disclinations pinned at the walls. Importantly, increasing the chirality favors a non-Newtonian response of the cholesteric witnessed by a sustained secondary flow. On the other hand, augmenting the shear-rate determines a decrease of the transition temperature, which leads the system either into the nematic state or, if the chirality is higher, into a blue-phase-like state. The dynamics is significantly different if an isotropic droplet is embedded in the sample. If the interface anchoring is weak, we generally find a reduction of the secondary flow with respect to the droplet-free cholesteric sample, an effect even more pronounced if the interface anchoring is strong. We ascribe the latter effect
to the resistance encountered by the flow to overcome a larger elastic deformation of the director field, in particular near the droplet interface where two fully in-plane topological defects form. These results are overall confirmed when the chirality is increased, although here an intense secondary flow, comparable to that observed in the droplet-free sample, is found for low shear-rate and weak anchoring. This highlights the fact that the rheological response displayed by an inverted cholesteric droplet has a complex landscape where a key role is played by at least three quantities: Shear rate, elasticity, and chirality. 

Our results represent a first step on the study of the rheology of an inverted cholesteric droplet, and are of possible interest for designing CLC based devices built from an emulsion. This opens up several directions for future research. It would be worth investigating the case in which tangential anchoring is considered (both at the droplet interface and at the walls) in a cholesteric sample whose axis is perpendicular to the walls. Besides yielding to the formation of different topological defects (such as a twisted Saturn ring in 3D), such configuration is expected to display a rich dynamical behavior like that observed with homeotropic interface anchoring which strongly depends upon the cholesteric pitch (or the chirality). This can pave the way to extend the study to include several droplets, either in a monodisperse setup or in the more intriguing polydisperse one where different droplet interface anchoring sets may be considered. One can envisage, for instance, the design of a new soft material made up of highly-packed isotropic droplets whose resistance to deformation could be sustained by the liquid crystal dispersed in between. Finally, although three-dimensional simulations can be computationally demanding, it would be surely of great interest to investigate the physics of more realistic systems in order, for instance, to minimize finite size effects. However we note that quasi two-dimensional liquid crystal devices could be experimentally realized, such as that described in Ref.~\cite{cluzeau}, where a smectic-C film surrounding droplets, obtained by nucleation, is proposed.

\vspace{6pt}

\authorcontributions{F. F., G. G., A. L., E. O. and A. T. designed and performed the research and wrote the manuscript.}

\conflictsofinterest{``The authors declare no conflict of interest.''}

\reftitle{References}

\end{document}